\documentclass[%
 reprint,
 amsmath,amssymb,
 aps,
]{revtex4-2}

\usepackage{graphicx}
\usepackage{dcolumn}
\usepackage{bm}
\usepackage{color}
\usepackage{braket}
\usepackage{booktabs}
\usepackage{comment}
\usepackage{aas_macros}

\begin{document}


\title{Cosmic variance or galaxy bias? Disentangling finite-volume \\ and galaxy formation effects in cosmological analysis}

\author{Francesco Sinigaglia$^{1,2,3,4}$}
\email{IFPU fellow: fsinigag@sissa.it}
\author{Francisco-Shu Kitaura$^{5,6}$}
\vspace{0.5cm}
\affiliation{
$^{1}$Institute for Fundamental Physics of the Universe (IFPU), Via Beirut 2, I-34151 Trieste, Italy}
\vspace{0.2cm}
\affiliation{
$^{2}$SISSA - International School for Advanced Studies, Via Bonomea 265, 34136 Trieste, Italy}
\vspace{0.2cm}
\affiliation{
$^{3}$INAF - Osservatorio Astronomico di Trieste, Via G. B. Tiepolo 11, I-34131 Trieste, Italy}
\vspace{0.2cm}
\affiliation{
$^{4}$INFN – National Institute for Nuclear Physics, Via Valerio 2, I-34127 Trieste, Italy}
\vspace{0.2cm}
\affiliation{
$^{5}$Instituto de Astrof\'isica de Canarias, Calle via L\'actea s/n, E-38205, La  Laguna, Tenerife, Spain
}
\vspace{0.2cm}
\affiliation{
$^{6}$Departamento  de  Astrof\'isica, Universidad de La Laguna,  E-38206, La Laguna, Tenerife, Spain}

\date{\today}

\begin{abstract}
Current and forthcoming galaxy redshift surveys, such as DESI and Euclid, are going to bring cosmological analysis to an unprecedentedly exquisite level of precision in the determination of the cosmological parameters. However, these efforts require a high degree of control over theory and systematics, to obtain unbiased results. In this sense, the cosmic variance associated to finite-volume effects represents a major challenge and should adequately accounted for. In this work, we revisit the definition of cosmic variance and develop a novel framework to describe it using a `galaxy biasing' formalism. In particular, we use halo/galaxy Eulerian perturbation theory to relate the density field from an arbitrary cosmic realization to its counterpart having statistical properties reproducing the ensemble average, introducing a new set of bias parameters. We then apply this idea to the description of the non-linear shift of BAO, disentangling the source of uncertainty from cosmic variance and galaxy biasing associated with the measurement of the BAO scale. We finally check our analytical argument against cosmological variance-suppressed $N$-body simulations, finding an expected reduction in the uncertainty on the BAO peak position. We conclude that extra care should be used when inferring cosmological information from perturbative approaches involving the estimation of bias parameters and propose new practical strategies to optimally leverage the novel formulation of cosmic variance presented herein in cosmological analysis.

\end{abstract}

\maketitle

\section{Introduction}\label{sec:intro}

Current and forthcoming large-scale galaxy redshift surveys, such as DESI \cite{Levi2013} and Euclid \cite{Amendola2018}, are going to deliver unprecedented three-dimensional maps of the Universe, whose analyses promise to shed light onto unsolved key questions about cosmology and fundamental physics, such as unveiling the nature of dark matter \citep[e.g.,][]{Braglia2025,Kumar2025} and dark energy \citep[e.g.,][]{Cortes2024,Li2024,Zheng2024,UrenaLopez2025,Lodha2025}, testing the validity of General Relativity as theory of gravity \citep[e.g.,][]{Beutler2020,Bonvin2023,Lepori2025}, and probing primordial non-Gaussianities and the inflationary scenario \citep[e.g.,][]{BermejoCliment2025,Chaussidon2025}, among others. In particular, the DESI collaboration has recently presented the strongest detection ever of Baryon Acoustic Oscillations (hereafter BAO), achieving subpercent precision in the determination of the position of the acoustic peak \cite{DESIDRII2025,DESIDRIILya2025}. 

In this golden era of precision cosmology, it is very timely to develop a complete understanding of all the ingredients associated to the measurements of cosmological parameters and, in particular, their related uncertainties. In this work, we analyze two important uncertainty contributions, which are strongly coupled: (i) cosmic variance, arising from the intrinsic variance imprinted in the initial density field, and (ii) galaxy biasing, describing galaxy formation effects. The former stems from the particular random realization of the Gaussian random field. The latter concerns the complex non-linear processes driving galaxy formation under different physical conditions, and depends on the characteristics of the host halo and of the cosmic web environment where it resides, among others. Theoretical analytical models typically include a set of free parameters aimed at describing galaxy biasing and fit them to observations. However, in this practical framework, it is impossible to know the relative contribution of galaxy bias and cosmic variance and the latter is not taken into account at the level of the estimation of the bias parameters, but only as an additional uncertainty contribution. In other words, the estimation of the bias parameters is performed by assuming the average matter power spectrum, and neglecting the cosmic variance effect coming from surveying a limited cosmological volume. 

In this paper we aim at disentangling cosmic variance and galaxy biasing and estimate separately the contributions coming from these two effects. To do so, we present a novel formalism based on halo/galaxy Eulerian perturbation theory, to treat cosmic variance in a similar way as done for galaxy biasing, and mathematically disentangle the two effects. This is particularly relevant for several applications, such as e.g. (i) BAO reconstruction, in which one needs to make assumptions on the numerical value for the galaxy bias, (ii) cosmological information extraction, such as inferring cosmological parameters from BAO, and (iii) galaxy evolution studies which use the galaxy bias as observational constraints to shed light onto the physics driving galaxy formation.  

The paper is structured as follows. In Sect.~\ref{sec:eul_pt}, we briefly review the Eulerian perturbation theory approach. In Sect.~\ref{sec:cosmic_variance}, we present the formalism to model cosmic variance underlying this work. In Sect.~\ref{sec:ic_ssc}, we discuss the integral constraint and incorporate it in the cosmic variance treatment presented herein. In Sect.~\ref{sec:bao_shift}, we apply the novel cosmic variance formulation to the non-linear BAO shift. Sect.~\ref{sec:sims} presents the numerical results validating the analytical framework. In Sect.~\ref{sec:surveys}, we present a few strategies to practically apply the ideas presented in this work to galaxy surveys and optimally extract cosmological information. We conclude in Sect.~\ref{sec:conclusions}.


\section{Eulerian perturbation theory}\label{sec:eul_pt}

According to \cite{Bernardeau2002}, the expansion of the matter density
contrast in Eulerian perturbation theory (EPT) can be written as a sum of
perturbative contributions,
\begin{equation}
\delta(\mathbf{k},\eta)
=
\sum_{n=1}^{\infty} \delta^{(n)}(\mathbf{k},\eta),
\end{equation}
where $\delta^{(n)}$ denotes the $n$-th order contribution and $\eta$ is the
time variable (e.g.\ conformal time or $\ln a$).

If we discretize Fourier modes (or grid cells) and collect the density field
$\delta(\mathbf{k})$ into a vector with indices $i,j,k,\ldots$, the convolution
integrals appearing in perturbation theory can be written as discrete sums
with coupling matrices (or tensors). The non-linear density field then takes
the form
\begin{equation}
\delta_i
=
\delta_i^{\mathrm{L}}
+
\sum_{j k} F^{(2)}_{i j k}\,
\delta_j^{\mathrm{L}}\,\delta_k^{\mathrm{L}}
+
\sum_{j k l} F^{(3)}_{i j k l}\,
\delta_j^{\mathrm{L}}\,\delta_k^{\mathrm{L}}\,\delta_l^{\mathrm{L}}
+\cdots ,
\end{equation}
where $\delta^{\mathrm{L}}$ denotes the linear density field and
$F^{(n)}_{i j_1 \cdots j_n}$ are the discretized Eulerian perturbation theory
kernels encoding mode coupling (for details see, e.g., \cite{Bernardeau2002}.

In the continuum limit, the $n$-th order contribution in Fourier space can be
written explicitly as
\begin{equation}
\begin{split}
\delta^{(n)}
(\mathbf{k},\eta)
&=
D^n(\eta)
\int \frac{d^3 q_1}{(2\pi)^3}
\cdots
\frac{d^3 q_n}{(2\pi)^3}
\,
(2\pi)^3
\\ & \delta_D
\!\left(
\mathbf{k}
-
\sum_{a=1}^{n} \mathbf{q}_a
\right)
F_n(\mathbf{q}_1,\ldots,\mathbf{q}_n)
\prod_{a=1}^{n} \delta_L(\mathbf{q}_a) \, ,
\end{split}
\end{equation}
where $D(\eta)$ is the linear growth factor, $\delta_D$ denotes the Dirac delta
distribution enforcing momentum conservation, and $F_n$ are the (symmetrized)
Eulerian perturbation theory kernels.

Collecting the perturbative contributions and transforming back to real space,
the non-linear matter density contrast can be written schematically as
\begin{equation}
\delta
(\mathbf{x})
=
\delta_L(\mathbf{x})
+
\delta^{(2)}(\mathbf{x})
+
\delta^{(3)}(\mathbf{x})
+
\cdots ,
\end{equation}
At second order, the density field admits the operator expansion

\begin{equation}
\delta^{(2)}(\mathbf{x})
=
c_1\,\delta_L^2(\mathbf{x})
+
c_2\,s_{ij}(\mathbf{x})\,s^{ij}(\mathbf{x})
+
c_3\,\delta_L(\mathbf{x})\,\nabla^2 \Phi(\mathbf{x}),
\end{equation}
where the coefficients $c_1$, $c_2$, and $c_3$ are fixed by the
underlying gravitational dynamics (e.g.\ $c_1 = 17/21$ and $c_2 = 2/7$ in an
Einstein--de Sitter universe),
 $s_{ij}$ is the tidal tensor,
\begin{equation}
s_{ij}(\mathbf{x})
=
\left(
\frac{\partial_i \partial_j}{\nabla^2}
-
\frac{1}{3}\delta_{ij}
\right)\delta_L(\mathbf{x}) \, .
\end{equation}

Neglecting total-derivative and convective terms, or equivalently working with
locally smoothed fields, the non-linear density contrast may be written in the
schematic form
\begin{equation}
\delta
(\mathbf{x})
=
\delta_L(\mathbf{x})
+
c_1\,\delta_L^2(\mathbf{x})
+
c_2\,s_{ij}(\mathbf{x})\,s^{ij}(\mathbf{x})
+
\mathcal{O}(\delta_L^3)
\end{equation}
This expression makes it explicit that, although gravitational evolution is
intrinsically non-local, Eulerian perturbation theory necessarily generates
local composite operators such as $\delta_L^2(\mathbf{x})$, accompanied by
non-local tidal contributions. In the remainder of the paper, we neglect for simplicity non-local terms, and use only the following local relation
\begin{equation}\label{eq:dm_pt}
\delta
(\mathbf{x})
=
\delta_L(\mathbf{x})
+
c_1\,\delta_L^2(\mathbf{x})
+ c_4\,\delta_L^3(\mathbf{x}) + \dots
\end{equation}

Afterwards, on sufficiently large scales where the density perturbations are small, the galaxy density can be modelled in a similar perturbative way as powers of the non-linear density field through the series:
\begin{equation}\label{eq:gal_pt}
\delta_g
(\mathbf{x})
= b_0 +
b_1\delta(\mathbf{x})
+
\frac{1}{2}b_2\,\delta^2(\mathbf{x}) + \dots .
\end{equation}
We notice that we will refer to `galaxy bias' throughout the paper, but the same argument holds true for any biased tracer of the dark matter field. This expression provides a practical framework to describe the clustering of dark matter tracers by means of perturbation theory techniques. In this framework, the two-point correlation function of the dark matter field can computed as $\xi_{\rm}=\langle \delta({\bf x_1})\delta({\bf x_2}))$, and similarly for galaxies $\xi_{\rm g}=\langle \delta_{\rm g}({\bf x_1}) \delta_{\rm g}({\bf x_2}) \rangle$.  While for this paper we have limited the Eulerian expression to just a local relation for simplicity, it is a well-established results that the galaxy bias has crucial non-local contributions up to third order \citep[e.g.,][]{McDonalRoy2009,Werner2020,Kitaura2022}. This simple formalism sets the basis for our treatment and will be used in the remainder of the paper.

From Eq. \ref{eq:gal_pt}, it is evident that the clustering and phenomenology of the biased field is determined by the parameters $b_1$, $b_2$, and others used to describe the model. These value for these parameters depend on redshift, on the analyzed tracer and its properties (e.g., the bias parameters will be different for different galaxy populations), and are meant to model all the physical processes related to the considered tracers: halo/galaxy formation and its dependence on the large-scale structure environment and on the anisotropic clustering, galaxy evolution, and baryon physics, among others. 


\section{A novel framework for cosmic variance}\label{sec:cosmic_variance}

The Eulerian framework developed in Sect. \ref{sec:eul_pt} implicitly assumes an infinite volume and provides analytical expressions corresponding to the ensemble average of the dark matter and halo/galaxy fields. How does this picture change when we consider an arbitrary realization of the Universe in a finite volume? The statistical properties of such a realization will tend to converge to the ensemble average on small scales, but they may diverge significantly at the largest scales due to their poor sampling. This is a well-known situation for the large-scale modes of the power spectrum and analogously for the two-point correlations functions measurements on large scales. While this issue is normally treated as just a source of uncertainty, it has a deeper physical meaning: in fact, whenever a perturbative framework is applied to perform cosmological measurements from a clustering observables, the found bias parameters $b_1$ and $b_2$ will include the degenerate effect of cosmic variance and of galaxy biasing. This aspect may become a problem in several aspects of cosmological analysis, in which a value for the bias needs to be assumed, as is the case for BAO reconstruction. 

To disentangle the two aforementioned effects, we introduce in this work a novel way of treating the density field from arbitrary realization as a `biased' tracer of its counterpart version featuring statistical properties identical to the ensemble average, which is the quantity described in Eq. \ref{eq:dm_pt}. In particular, by calling $\delta'$ the density field from an arbitrary realization in a finite volume, we can write:
\begin{equation} \label{eq:biased_dm}
\delta^\prime
(\mathbf{x})
= b_{0,{\rm dm}} + 
b_{1,{\rm dm}}\delta(\mathbf{x})
+
\frac{1}{2}b_{2,{\rm dm}}\,\delta^2(\mathbf{x}) + \dots \quad .
\end{equation}

This expression assumes again that the `biased' dark matter field has just tiny fluctuations with respect to its counterpart with average properties. In the limit for the volume which tends to infinity, $b_{\rm 1,dm}\rightarrow1$, $b_{\rm 2,dm}\rightarrow0$, and $\delta'\rightarrow\delta$.

In this picture, we have disentangled the effects of cosmic variance and of galaxy biasing, at the expense of introducing a new set of parameters modelling the former. In this way, the bias parameters will now be variance-free and will truly correspond to the biasing effect. 

To develop an intuitive understanding of the meaning of $b_{\rm 1,dm}$ and $b_{\rm 2,dm}$, let us use the formalism of Lagrangian perturbation theory (LPT). In LPT, the Lagrangian and Eulerian coordinates ${\bf q}$ and ${\bf x}$ are related through the displacement field ${\bf \Psi}$ as: ${\bf x}={\bf q}+{\bf \Psi}({\bf q})$. The displacement field in first-order LPT (Zel'dovich approximation \cite{Zeldovich1970}) is related to the initial linear density field through the relation
\begin{equation}\label{eq:disp}
    \delta_l \sim -\nabla\cdot{\bf \Psi}({\bf q}) \quad .
\end{equation}
Given that the cosmic variance stems from the initial conditions, one can write an expression similar to Eq. \ref{eq:biased_dm} also for the initial density field from an arbitrary realization. In this sense, the two bias parameters $b_{\rm 1,l}$ and $b_{\rm 2,l}$ will control the displacement field through Eq. \ref{eq:disp}, and hence, determine the non-linear dynamics of the system. The values of the initial density field are fully-characterized mode by mode in Fourier space by the amplitude --- regulated by the linear power spectrum --- and the phase. Both the amplitude and the phase contribute to cosmic variance. It is straightforward to understand how the amplitude, by definition, sources cosmic variance in the two-point statistics. The phases, instead, do not necessarily impact two-point statistics, but induce cosmic variance in the one-point statistics. This is clear if one considers paired simulations, whose initial density values are related through a rotation of angle $\theta\rightarrow\theta+\pi$, i.e. a change of sign in configuration space. In this case, such simulations will have exactly the same amplitude mode by mode, but the one-point PDF will be symmetric with respect to zero density. These effect in the initial conditions are then propagated in a nontrivial way through the non-linear evolution of cosmic structures.       

\section{Integral constraint}\label{sec:ic_ssc}

In large scale structure cosmology, another source of uncertainty stemming from surveying a finite volume is the so called {\it integral constraint}, which consists in a potential bias arising from the estimation of the mean density of the Universe using the surveyed volume itself. In particular, inside a survey volume the estimator forces the average density inside the observed volume to vanish:
\begin{equation}
    \int_V \delta (\mathbf{x})d^3\mathbf{x}=0 \quad .
\end{equation}
However, the mean overdensity $\delta_V$ within the survey volume does not necessarily vanish:
\begin{equation}
    \delta_V \equiv \int_V \delta (\mathbf{x})d^3\mathbf{x}\neq 0 \quad .
\end{equation}
Therefore, the measured density fluctuations are actually $\hat{\delta}(\mathbf{x})=\delta(\mathbf{x})-\delta_V$. In this picture, the largest-scale modes are suppressed,
the clustering amplitude is biased low on large scales,
and the measured correlation function differs from the true one.

Therefore, cosmic variance quantifies the statistical fluctuations of the survey mean due to long modes, while the integral constraint represents the bias which is introduced by forcing those long modes to zero by the estimator. In this sense, the integral constraint is, strictly speaking, not a cosmic variance term, but a distortion that arises from the estimator due to the cosmic variance. While in a standard analysis the integral constraint is treated by modifying the theory at the clustering level, in this formalism we simply express the final observed galaxy density field by using Eq.~\ref{eq:gal_pt}:
\begin{equation}
    \hat{\delta_g}(\mathbf{x}) = \delta_g(\mathbf{x})  - \delta_{g,V} \quad ,
\end{equation}
where $\delta_{g,V}$ is the measured mean galaxy number density within the survey volume. By expanding Eq.~\ref{eq:gal_pt} and defining $b_0^\prime=b_0 - \delta_v$, the contribution from the integral constraint get reabsorbed in the definition of the constant term $b_0^\prime$, therefore it implies no extra bias parameters. In this way, the cosmic variance contribution can be disentangled from galaxy bias as explained in the previous sections with not further complications.

\section{The non-linear BAO shift}\label{sec:bao_shift}

In this section, we apply the model for cosmic variance introduced in Sect. \ref{sec:cosmic_variance} to study its impact on the determination of the non-linear BAO scale from clustering measurements. In particular, we rely on the formalism by \cite{SherwinZaldarriaga2012}, wherein the authors derive analytical expressions for the non-linear BAO shift of the density field and of biased tracers by means of Eulerian perturbation theory. Let us express the density field as $\delta({\bf x})=\delta_{l}({\bf x})+\delta_{s}({\bf x})$, i.e. as sum of a long-wavelenght component $\delta_l$ --- sourcing the BAO shift --- and a short-wavelength component $\delta_s$ on scales comparable to the BAO scale. In this framework, the non-linear BAO shift of the real space correlation function for the dark matter field reads \cite{SherwinZaldarriaga2012}:
\begin{equation}
    \Delta\alpha=\alpha-1=\frac{131}{105}\langle\delta_l^2 \rangle
\end{equation}
and for biased tracers:
\begin{equation}\label{eq:bao_shift_biased}
    \Delta\alpha=\alpha-1=\frac{131}{105}\left(1 + \frac{70}{131} \frac{b_2}{b_1}\right) \langle\delta_l^2 \rangle \quad ,
\end{equation}
where $b_1$ and $b_2$ are the Eulerian bias parameters defined in Eq. \ref{eq:gal_pt}. At this stage, $b_1$ and $b_2$ include the degenerate effect from true galaxy biasing and from cosmic variance. To disentangle the two, let us rewrite Eq. \ref{eq:bao_shift_biased} in two parts: one part related to the intrinsic cosmic variance of the dark matter field with respect to the ensemble average, controlled by the bias parameters $b_{1,{\rm dm}}$ and $b_{2,{\rm dm}}$
\begin{equation}\label{eq:bao_shift_dm}
   \Delta\alpha_{\rm dm}=\frac{131}{105}\left(1 + \frac{70}{131} \frac{b_{2,{\rm dm}}}{b_{1,{\rm dm}}}\right) \langle\delta_l^2 \rangle \quad 
\end{equation}
and the other related to galaxy biasing and controlled by the parameters $b_{1,{\rm g}}$ and $b_{2,{\rm g}}$
\begin{equation}\label{eq:bao_shift_gal}
   \Delta\alpha_{\rm g}=\frac{131}{105}\left(1 + \frac{70}{131} \frac{b_{2,{\rm g}}}{b_{1,{\rm g}}}\right) \langle\delta_l^{\prime2} \rangle \quad ,
\end{equation}
were we have expressed also the `biased' dark matter field as $\delta^\prime({\bf x})=\delta_{l}^\prime({\bf x})+\delta^\prime_{s}({\bf x})$. We notice that in Eq.~\ref{eq:bao_shift_dm} we are considering only the long-range component $\delta_l$ and neglecting the short-wavelength part, which has negligible contribution to the BAO shift \cite{SherwinZaldarriaga2012}. In this formulation, it is straightforward to understand that cosmic variance alone can induce an additional non-negligible BAO shift, especially in small volumes, which can be comparable to the one induced by galaxy biasing. Therefore, one should pay extra care when estimating the BAO shift using perturbative approaches, such as e.g. Effective Field Theory, applied to small volumes. These results can be intuitively understood by considering that the BAO shift is driven at first order by the Zel'dovich displacement \cite{Zeldovich1970}, which will be sensitive to an excess overdensity in a simulation volume with respect to the ensemble average, and will typically result in an extra shrinking of the BAO scale compared to the mean BAO shift. It is important to stress that in a realistic cosmological analysis, the estimation of the bias parameters is performed by relating the `ensemble average' dark matter field to the biased tracers of a specific realizations. Therefore, by construction the estimated bias parameters will include the degenerate contribution from the bias and from cosmic variance. While this is in principle accounted for in the estimation of the uncertainty of the bias parameters, in Sect. \ref{sec:surveys} we propose a few strategies to disentangle these effects and suppress the contribution to the uncertainty on the estimated cosmological parameters due to cosmic variance.


\section{Empirical validation with simulations}\label{sec:sims}

In this section, we investigate the variance associated to the measurements of the BAO scale by using N-body simulations, with the aim of showing that controlling the variance in the initial conditions results in a suppression of the uncertainty on the BAO scale. To this end, we consider the \texttt{UNIT} simulations suite \cite{Chuang2019}, designed to provide precise predictions for non-linear statistics of the galaxy distribution. The simulations of the \texttt{UNIT} project were run either with the standard \texttt{GADGET} TreePM code \cite{Springel2001,Springel2005} or with the particle mesh \texttt{FastPM} code \cite{Feng2016}, and include a variety of different realizations in different volumes and resolutions. Specifically, the flagship runs, as well as other simulation within the suite, were run by adopting the fixing and pairing techniques for initial conditions, to suppress cosmic variance and provide a robust framework for galaxy clustering analysis.  

In this work, we consider the suite of $2\times100$ fixed-and-paired simulations run with \texttt{FastPM} in $V=(1~h^{-1}{\rm Gpc})^3$ volumes and with $N=1024^3$ dark matter particles. In particular, we consider the output at $z=0$ and make use of the measurements of the isotropic two-point correlation function in real space for the dark matter particles and both in real and in redshift space for the haloes made available by the \texttt{UNIT} consortium. 

We estimate the BAO peak position in a model-independent fashion as follows. We first fit a broad-band shape polynomial of the type $A(r)=ar^{-2}+br^{-1}+c$, with $a,b,c$ free parameters \citep[see e.g.,][]{Xu2012,VargasMagan2014}. Afterwards, we subtract it from the full correlation function, to keep only the BAO peak `wiggly' part. Then, we fit a Gaussian function to the peak component of the correlation function, in the scales range $r\in [90,120]~h^{-1}{\rm Mpc}$, and consider the centroid of the Gaussian as a model-independent position of the acoustic peak. We notice that more sophisticated ways of fitting the BAO peak position based on templates of the input linear power spectrum are available from the literature \cite[e.g.,][and references therein]{Zhao2020}. However, since here we are only interested in a model-independent estimate of the position of the acoustic peak and not in determining the $\alpha$ parameter, we limit our methodology to what described above. Finally, we build the distribution of BAO peak positions for each of the studied cases mentioned above and estimate the ensemble average value as the median of the distributions and use the $(84th-50th,50th-16th)$ percentiles as associated upper and lower uncertainties. 

We apply this BAO fitting procedure on the following sets of simulations:
\begin{itemize}
\item $100$ standard simulations;
\item $100$ fixed-amplitude simulations;
\item $2\times50$ paired simulations;
\item $2\times 50$ fixed-and-paired simulations.
\end{itemize}

\begin{figure*}
    \centering
    \includegraphics[width=0.49\textwidth]{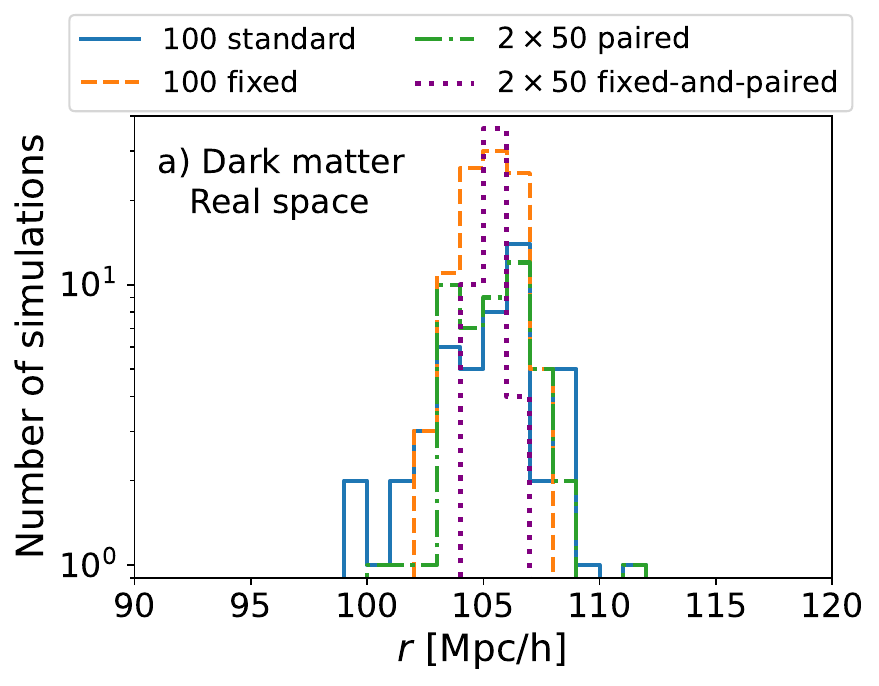}
    \includegraphics[width=0.49\textwidth]{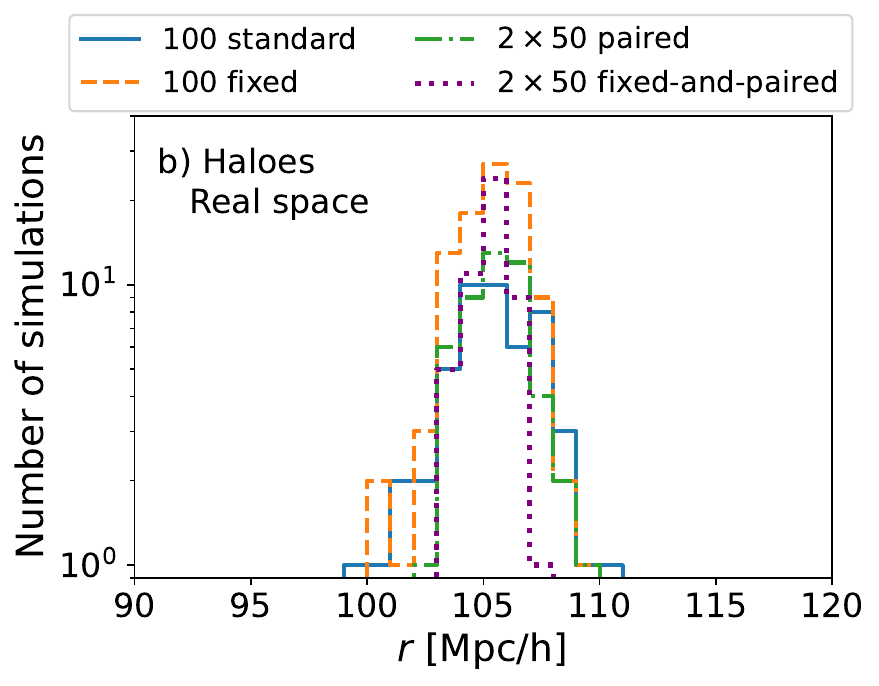}
    \includegraphics[width=0.49\textwidth]{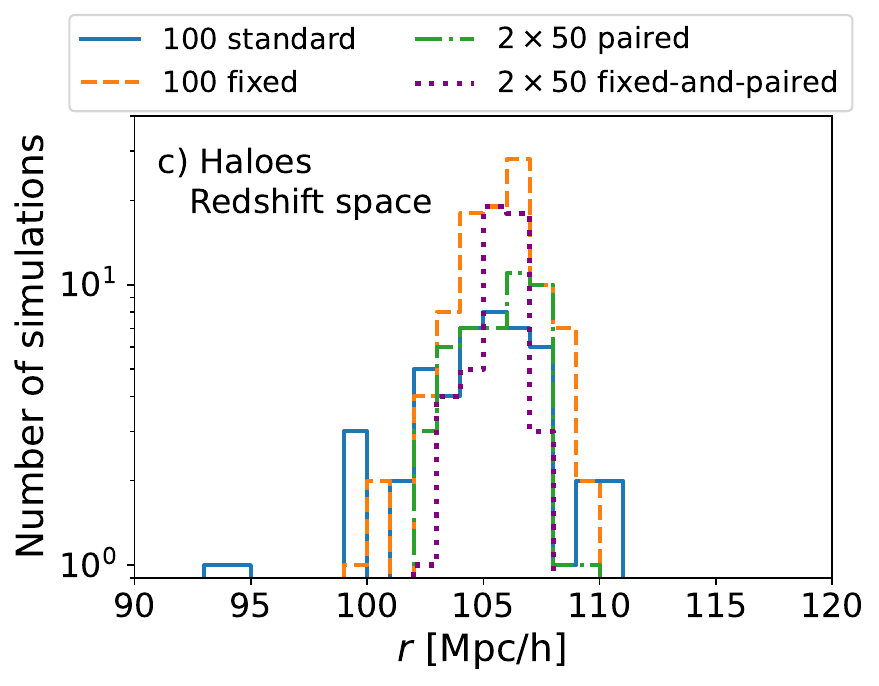}
    \caption{Distributions of the BAO peak position measured from the two-point correlation function from \texttt{FastPM} simulation from the \texttt{UNIT} project, as measured from the dark matter field in real space (a), the halo field in real space (b) and in redshift space (c) at $z=0$. We display the distribution from $100$ standard simulations as a blue solid line, from $100$ fixed-amplitude simulations as an orange dashed line, from $2\times 50$ paired realizations as a green dotted-dashed line, and from $2\times 50$ fixed-and-paired simulations as a purple dotted line.}
    \label{fig:bao_hist}
\end{figure*}

We show the resulting BAO peak position distributions in Fig. \ref{fig:bao_hist} and report the numerical results in Table \ref{tab:bao_results}. We notice that in all the studied case, the BAO peak position is well-converged at subpercent level in the different studied cases and there is no apparent systematics introduced by using fixing and/or pairing techniques with respect to the ensemble of the $100$ standard simulations. The sample of $100$ fixed-and-paired simulations has always the smallest variance among the studied cases, by a factor $\sim 5.4$ with respect to the traditional ensemble average of the dark matter in real space, by a factor $\sim 3.4$ for the haloes in real space, and by a factor $\sim 3.9$ for the haloes in redshift space. The fixing and pairing techniques produce a suppression in the uncertainty on the BAO peak: a factor $\sim 2.1$ (fixing) and $\sim 1.5$ (pairing) for the real-space dark matter field, a factor $\sim 1.6$ (fixing) and $\sim 1.7$ (pairing) for the real-space halo field, and a factor $\sim 1.8$ (fixing) and $\sim 1.7$ (pairing) for the redshift-space halo field. The same results can be also visually appreciated by looking at the distributions shown in Fig. \ref{fig:bao_hist}. From these results, we infer that fixing and pairing have different, complementary effects on the overall error budget on the BAO peak position: the fixing technique reduces the large-scale bias of each individual realization with respect to the ensemble average, while stacking couples of paired simulations reduced the uncertainty due to the deviation from the average non-linear shift. 

\begin{table}[]
    \centering
    \vspace{0.3cm}
    \begin{tabular}{lcccc}
    \toprule
    {\bf Sample}  & & $\mathbf{r_B}$ & $\mathbf{\Delta r_{B,{\rm up}}}$ & $\mathbf{\Delta r_{B,{\rm low}}}$\\
    \midrule
     {\bf DM real space} & \hspace{0.3cm} & & & \\
    \midrule
    $100$ standard &  & 105.6 & 2.4 & 2.6  \\
    $100$ fixed &  & 105.3 & 1.3 & 1.3  \\
    $100$ paired &  & 105.5 & 1.6 & 1.9  \\
    $100$ fixed-and-paired &  & 105.4 & 0.5 & 0.5  \\
    \midrule
    \midrule
     {\bf Haloes real space} & & & & \\
    \midrule
    $100$ standard &  & 105.3 & 2.6 & 2.0  \\
    $100$ fixed &  & 105.4 & 1.3 & 1.6  \\
    $100$ paired &  & 105.8 & 1.4 & 1.4  \\
    $100$ fixed-and-paired &  & 105.4 & 0.6 & 0.7  \\
    \midrule
    \midrule
     {\bf Haloes redshift space} & & & & \\
     \midrule
     $100$ standard &  & 105.8 & 3.8 & 2.8  \\
    $100$ fixed &  & 105.9 & 1.7 & 1.8  \\
    $100$ paired &  & 105.9 & 1.8 & 2.1  \\
    $100$ fixed-and-paired &  & 105.8 & 0.7 & 0.9  \\
    \bottomrule
    \end{tabular}
    \caption{Numerical results from the fitting of the BAO scale in different samples of $N$-body simulations from the \texttt{UNIT} project. The first column summarizes the investigated sample of simulations, the second column reports the median BAO scale value from the full distribution of the sample, and the third and fourth column report the upper and lower uncertainty values associated to the BAO peak position, estimated as ($84th-50th,50th-16th$) percentiles. The three sections of the Table refer to the results for the real-space dark matter field (top), the real-space halo field (mid), and the redshift-space halo field (bottom) at $z=0$.}
    \label{tab:bao_results}
\end{table}

\section{Application to galaxy surveys}\label{sec:surveys}

In this section, we briefly illustrate two potential different ways stemming from the idea presented in this work, to separate the contributions from cosmic variance and galaxy biasing in BAO analysis. In particular, this could be achieved through:
\begin{itemize}
\item {\it An optimal combination of the clustering from different complementary tracers}. It has been shown that adequately fitting the  post-reconstruction combined BAO signal from different tracers, such as e.g. galaxies and cosmic voids, can lead up to a $\sim10-20\%$ precision improvement in the determination of the position of the acoustic peak \citep[e.g.,][]{Zhao2020,Zhao2022}. Here, we propose to shift the focus and tailor the combination in such a way not to necessarily maximize the achieved statistical precision, but rather to cancel out cosmic variance. This can be achieved by (i) revisiting the definition of the two pre-reconstruction samples in such a way that they trace exactly complementary density regimes, or (ii) finding the optimal set of weights in the computation of two-point correlation function from the combined post-reconstruction samples. This strategy aims at mimicking the usage of two paired realizations, fully exploiting the information from the full density range and should be accurately tested on detailed mock catalogs; 

\item {\it The application of cosmic variance suppression techniques (e.g. fixing and pairing techniques) both to the true observed data and to the mock catalogs adopted to estimate the covariance matrices.} In particular, the application of backward inversion methods relying on the negative displacement \citep[see e.g.,][and references therein]{Schmittful2017,HadaEisenstein2018} field or forward Bayesian reconstruction techniques  \citep[e.g.,][and references therein]{Kitaura2013,Jasche2013,Wang2013,Jasche2019,Kitaura2021,Rossello2025} to obtain the initial conditions underlying the observed data, coupled to fixing and pairing techniques, would yield a variance-suppressed version of the observed Universe and its (paired) anti-Universe. While fixed-and-paired mock catalogs feature a significant suppression of cosmic variance and are therefore not suited to be employed in traditional cosmological analysis \cite{Klypin2020}, we argue that their usage would be consistent and legitimate if applied to the analysis of `fixed-and-paired' observed clustering measurements. 
Extra care in this sense should be paid when performing the reconstruction, both in inversion and Bayesian forward models. In the former, the accuracy of the reconstruction method will significantly impact the final result when evolving the initial conditions forward in time after having applied the cosmic variance suppression techniques. As for the latter, one does not want to impose a given BAO shift through the cosmology adopted to compute the linear power spectrum, otherwise the final result will be naturally biased towards it. Potential ways of copying with this issue are sampling either the power spectrum, or the BAO shift as part of the reconstruction procedure. The former allows the practitioner to have a cosmology-agnostic reconstruction of the initial conditions, while the latter to obtain a posterior distribution for $\alpha$ to draw values from when generating samples for the initial conditions.   
\end{itemize}

We leave the exploration of these ideas for future publications.


\section{Conclusions}\label{sec:conclusions}

In this work, we have presented a novel formulation of cosmic variance in an Eulerian perturbation theory fashion. In particular, we have proposed a framework to study cosmic variance leveraging the formalism usually adopted to treat biased tracers of the dark matter field. Such formulation effectively disentangles the contribution of cosmic variance from the one of true galaxy bias by means of a new set of free parameters modelling the `cosmic variance bias' effect. We have applied this formalism to the perturbative approach to compute the non-linear BAO shift originally proposed by \cite{SherwinZaldarriaga2012}, showing that cosmic variance alone can account for a significant bias on the non-linear BAO shift, which is degenerate with the non-linear shift stemming from the non-linear gravitational growth of cosmic structures. 

Afterwards, we have tested the impact of cosmic variance on the determination of the BAO peak position from cosmological simulations. In particular, we considered a suite of simulations comprising $100$ standard, $100$ fixed, $2\times 50$ paired and $2\times 50$ fixed-and-paired simulations. We have determined the BAO peak position for each realization in a model-independent fashion and built the resulting distributions for each aforementioned class of simulations. In this way, we have shown that both the fixing and pairing techniques reduce the variance of the BAO scale distribution by a factor $\sim 1.5-2$ both for the dark matter field in real space and for the haloes in real and redshift space at $z=0$. This indicates that each of them removes one of the source of uncertainties, which are found to have a similar contribution to the total error budget. Specifically, the pairing technique averages out the shift from couples of simulations, while the fixing technique alleviates the contribution to the variance due to the large-scale bias effect. The combination of the two techniques yields a suppression of the variance by a factor $\sim 4-5$. 

Finally, we have proposed two practical strategies, to be tested in future work, to disentagle cosmic variance and galaxy biasing effects in a realistic cosmological analysis. Specifically, one strategy consists in optimally leveraging the complementary properties of different samples in a multitracer analysis, either through a proper definition of the samples or via a weighting scheme. Cosmic voids have been proposed as a complementary tracer for BAO measurements \cite{Kitaura2016,Liang2016,Zhao2016} and shown to improve galaxy-only BAO constraints when combined with galaxy clustering \cite{Zhao2020,Zhao2022}. These approaches could be naturally extended to a paired BAO analysis by optimizing the BAO scale shift between tracers, rather than the BAO signal-to-noise, which was the primary objective in those works. The more challenging possibility consists in applying variance and pairing suppression techniques to the reconstructed initial conditions used to generate paired constrained simulations of the Universe conditions underlying real observations, as well as to the initial conditions used to generate the mock catalogs used to estimate the covariance matrices.  To this end, we need to have a complex nonlinear and nonlocal bias description (see, e.g., \cite{Kitaura2022,Sinigaglia2024a, Coloma2024,Sinigaglia2024}) implemented in a full Bayesian framework \cite{Rossello2025}. In future work, we will test these techniques and assess the realistic gain in using them by applying this framework to DESI-like mock galaxy catalogs and combine it with state-of-the-art analysis techniques such as BAO reconstruction and Effective Field Theory. 
\vspace{0.5cm}
\begin{acknowledgments}
F.S. acknowledges the support from the IFPU Postdoctoral Fellowship scheme. F.S.K. acknowledge the Spanish Ministry of Economy and Competitiveness (MINECO) for financing the \texttt{Big Data of the Cosmic Web} project: PID2020-120612GB-I00/AEI/10.13039/501100011033. F.S. is also grateful to the {\it Instituto de Astrofísica de Canarias} for hospitality during the completion of this project.
\end{acknowledgments}

\appendix



\bibliography{main}

\end{document}